\documentclass[preprint,nofootinbib,amsmath,amssymb]{revtex4}

\usepackage[dvips]{graphicx}

\begin{document}

\title{Consistency Test of Dark Energy Models}

\author{Chien-Wen Chen}
\email{f90222025@ntu.edu.tw} %
\affiliation{Department of Physics, National Taiwan University, Taipei 10617, Taiwan, R.O.C.} %
\author{Je-An Gu}
\affiliation{Leung Center for Cosmology and Particle Astrophysics (LeCosPA),\\
National Taiwan University, Taipei 10617, Taiwan, R.O.C.
}

\author{Pisin Chen}
\affiliation{Kavli Institute for Particle Astrophysics and Cosmology,\\
SLAC National Accelerator Laboratory, Menlo Park, CA 94025, U.S.A. and\\
Department of Physics, Graduate Institute of Astrophysics and LeCosPA,\\
National Taiwan University, Taipei 10617, Taiwan, R.O.C.} %

\begin{abstract}
Recently we proposed a new approach to the testing of dark energy models based on the observational data. In that work we focused particularly on quintessence models for demonstration and invoked a widely used parametrization of the dark energy equation of state. In this paper we take
 the more recent SN Ia, CMB and BAO data, invoke the same parametrization,
 and apply this method of consistency test to five categories of dark
 energy models, including the $\Lambda$CDM model, the generalized
 Chaplygin gas, and three quintessence models: exponential, power-law and
 inverse-exponential potentials. We find that the exponential potential
 of quintessence is ruled out at the $95.4\%$ confidence level, while the
 other four models are consistent with data. This consistency test can be
 efficiently performed since for all models it requires the constraint of
 only a single parameter space that by choice can be easily accessed.
\end{abstract}


\maketitle

\section{Introduction\label{sec:introduction}}
Compelling evidences from Type Ia supernovae (SN Ia) and other
cosmological observations show that the expansion of the universe is
undergoing an accelerating stage at late times (see
Ref.~\onlinecite{Frieman:2008sn} for reviews). Within the framework
of general relativity and assuming homogeneity and isotropy, this
indicates that there should exist an energy source, termed dark
energy, which provides a significant negative pressure to cause this
acceleration. Thus far the nature of dark energy remains unresolved
and is generally regarded as one of the most tantalizing problems in
cosmology. While a positive cosmological constant remains the
simplest realization of dark energy, current observational data have
not ruled out the possibility of a time-evolving dark
energy.\cite{Albrecht:2006um} As one awaits more information from
the future observations, the constraining power of the
next-generation observations and new analysis methods are being
pursued.\cite{Albrecht:2006um}$\textrm{--}$\cite{Zunckel:2008ti}

Many dark energy models have been proposed and studied. For the
cosmological constant, its value has been constrained by
observations (see Ref.~\onlinecite{Komatsu:2008hk}, for example).
The quintessence model, which invokes a time-varying scalar
field,\cite{Caldwell:1998ii}$\textrm{--}$\cite{Boyle:2001du}
generally allows its energy density and equation of state to evolve
with time. There are various quintessence models with different
potential forms (see,
Refs.~\onlinecite{Frieman:1995pm}$\textrm{--}$\onlinecite{Steinhardt:1999nw},~\onlinecite{Bozek:2007ti},
for example) that have been proposed. Studies of the
classification\cite{Caldwell:2005tm,Linder:2006sv} and the general
dynamical behavior\cite{Huterer:2006mv} of quintessence have been
carried out. There have been also works on the reconstruction of
quintessence
potentials\cite{Huterer:2006mv}$\textrm{--}$\cite{Sahlen:2006dn}$,$\cite{Gu:2008ch}
and the investigations on how future observational data can
constrain individual models of
quintessence\cite{Barnard:2007ta}$\textrm{--}$\cite{Bozek:2007ti}.
The generalized Chaplygin gas (see Ref.~\onlinecite{Bento:2002ps}
and references therein) has been proposed to either unify dark
matter and dark energy\cite{Bilic:2001cg} or to simply play the role
of dark energy\cite{Kamenshchik:2001cp,Sen:2005sk}. The constraint
of the generalized Chaplygin gas has been
obtained.\cite{Sen:2005sk}$\textrm{--}$\cite{Davis:2007na} While
these works have helped us study the possible nature of dark energy
and obtain constraints of the parameters of an individual dark
energy model, it should be desirable to explore possible means to
determine whether a particular dark energy model can be ruled out by
the observational data.

Recently, we introduced a new approach to testing the consistency
between a dark energy model and the observational
data.\cite{Gu:2008ch} We tested the exponential (see
Ref.~\onlinecite{Bozek:2007ti} and references therein) and the
power-law potentials\cite{Zlatev:1999tr,Steinhardt:1999nw} of
quintessence to demonstrate this method, which can be summarized as
follows. For each dark energy model, we look for a {\it
characteristic}, $Q(z)$, which in general can vary with the redshift
but is equivalent to a constant parameter within the domain of the
model. We further define the {\it measure of consistency},
$\mathcal{M}(z)$, as the derivative of $Q(z)$ with respect to the
redshift $z$. The observational data should allow a null value for
$\mathcal{M}(z)$ if the corresponding dark energy model is
consistent
 with them. If, however, the $\mathcal{M}(z)$ = 0 line lies outside certain confidence region, then that dark energy model is ruled out by the observational data
at the corresponding confidence level. To obtain the constraint on
the measure of consistency $\mathcal{M}(z)$ from the observational
data, a parametrization of the relevant physical quantity, such as
the equation of state or the luminosity distance, is required. We
have invoked a broadly used form of parametrization of the equation
of
state,\cite{Chevallier:2000qy,Linder:2002et}$,$\cite{Albrecht:2006um}
\begin{equation}
 w(z) = p_\textsc{de}(z) / \rho_\textsc{de}(z) = w_0 + w_a
(1-a) = w_0 + w_a z/(1+z) \,,  \label{eq:w0wa-parametrization}
\end{equation}
where $p_\textsc{de}(z)$ and $\rho_\textsc{de}(z)$ are the pressure
and the energy density of dark energy, respectively. The two
parameters in Eq.~(\ref{eq:w0wa-parametrization}) and the normalized
matter density at present, $\Omega_{m}$, define the parameter space,
$(w_0, w_a, \Omega_{m})$, through which the information from the
observational data can be extracted. A recent work that is close in
spirit to ours is that of Zunckel and Clarkson,\cite{Zunckel:2008ti}
who proposed consistency test of the cosmological constant via a
direct parametrization of the luminosity distance. In our
terminology, they use $Q(z) = 1- \rho_\textsc{de}(z)/\rho_{c}$,
which is equivalent to the constant $\Omega_m$ in the domain of the
cosmological constant, where $\rho_{c}$ is the critical density at
present. Sahni et al.\cite{Sahni:2008xx} also proposed null test of
the cosmological constant via the diagnostic $Om(z) =
(H^2(z)/H_0^2-1)/[(1+z)^3-1]$, which is equivalent to the constant
$\Omega_m$ in the domain of the cosmological constant, where $H(z)$
is the Hubble expansion rate and $H_0$ is the Hubble constant.

In this paper, we take the more recent data set and apply our method
of consistency test to five dark energy models, including the
cosmological constant, the generalized Chaplygin gas as the dark
energy component,\cite{Kamenshchik:2001cp,Sen:2005sk} and three
quintessence models: exponential, power-law and inverse-exponential
potentials\cite{Zlatev:1999tr,Steinhardt:1999nw}. The data set we
use includes a recently compiled ``Constitution set'' of SN Ia
data,\cite{Riess:2004nr}$\textrm{--}$\cite{Hicken:2009dk} the cosmic
microwave background (CMB) measurement from the five-year Wilkinson
Microwave Anisotropy Probe (WMAP) observation,\cite{Komatsu:2008hk}
and the baryon acoustic oscillation (BAO) measurement from the Sloan
Digital Sky Survey (SDSS)\cite{Eisenstein:2005su} and the $2$dF
Galaxy Redshift Survey ($2$dFGRS)\cite{Percival:2007yw}.

A conventional way to determine how well a dark energy model can fit
the observational data is the {\it model-based approach}, in which
one optimizes the parameters of each dark energy model based on the
observational data and then statistically assesses the goodness of
fit (see Ref.~\onlinecite{Davis:2007na}, for example). In such
approach one has to obtain the best fit for each set of parameters
specific to the particular dark energy model, which could be
tedious. In particular, in order to optimize the parameters of a
quintessence model one has to solve the field equation numerically
for each point in the parameter space, which can be computationally
intensive and time consuming.\cite{Abrahamse:2007te} In contrast, in
our approach we first constrain the parameters of the chosen
parametrization through the observational data, and then test
consistency of each dark energy model based on this set of
parameters. This can be more efficient than the model-based approach
when one deals with a large number of dark energy models. It is also
more direct and therefore much faster to constrain the parameter
space $(w_0, w_a, \Omega_{m})$ than to optimize the parameters of a
quintessence model in the model-based approach. The potential
downside of our method, however, would be that as long as one
invokes a specific form of parametrization, one might have
simultaneously imposed a prior, or bias, against certain dark energy
models. This issue requires a separate investigation and we are
currently pursuing that.\cite{Chen:2009} We note that the two
methods are different in spirit. The goodness of fit describes how
well a model can fit the observations. The consistency test, on the
other hand, examines whether the condition necessary for a model is
excluded by the observations. With in mind the pros and cons
mentioned above, we believe that the two methods, that is, the
model-based and ours, should be complimentary to each other in the
pursuit of revealing the nature of dark energy.

\section{Consistency Test of Dark Energy Models\label{sec:Consistency Test}}

\subsection{Formalism}  \label{sec:Formalism}
In this paper we consider a flat Friedmann-Lemaitre-Robertson-Walker (FLRW) universe and assume that it is dominated by pressureless matter and dark energy in the present epoch. The Hubble expansion rate, $H \equiv
\dot{a}/a$, is given by the Friedmann equations as
\begin{eqnarray}
H^2 (z) &=& \frac{8 \pi G_N}{3} \left[ \rho_{m}(z) +
\rho_\textsc{de}(z) \right] \nonumber \\
&=& H_0^2 \left[ \Omega_{m} (1+z)^3 + (1-\Omega_{m}) \exp
\left( 3 \int_{0}^{z} \left[ 1 + w(z') \right]
\frac{dz'}{1+z'} \right) \right] , %
\label{eq:H2(z)}
\end{eqnarray}
where the dark energy density
\begin{equation}
\rho_\textsc{de}(z) = \rho_c (1-\Omega_{m}) \exp \left( 3 \int_{0}^{z}
\left[ 1 + w(z') \right] \frac{dz'}{1+z'} \right) ,
\label{eq:rho phi (z)}
\end{equation}
and
\begin{equation}
\rho_c \equiv \frac{3H_0^2}{8 \pi G_N} \, . %
\label{eq:rho0}
\end{equation}
For quintessence as a dark energy model, the quintessence field and the potential are related to the equation of state, the Hubble expansion rate, and the dark energy density as follows \cite{Guo:2005ata,Gu:2008ch}.
\begin{equation}
\phi(z) - \phi_0 = \pm \int_0^z \frac{\sqrt{\left[ 1+w(z')
\right]\rho_\textsc{de}(z')}}{H(z')} \frac{dz'}{1+z'} \, , %
\label{eq:phi(z)}
\end{equation}
\begin{equation}
  V(z) = \left[ 1 - w(z) \right] \rho_\textsc{de}(z) / 2 \, . \label{eq:V(z)}
\end{equation}

We perform the consistency test of five dark energy models including the cosmological constant, the exponential potential, the power-law potential, the inverse-exponential potential, and the generalized Chaplygin gas as the dark energy component.

For the cosmological constant, the energy density $\rho_{\Lambda}$ is a constant. We define the characteristic $Q_\Lambda(z)$ as the dark energy density $\rho_\textsc{de}(z)$, which in general would evolve with the redshift but is equivalent to the constant parameter $\rho_{\Lambda}$ within the cosmological constant domain,
\begin{eqnarray}\label{eq:Qlambda}
 Q_\Lambda(z)
  & \equiv & \rho_\textsc{de}(z) \\
  & = & \rho_{\Lambda}\quad \textrm{for the cosmological constant.}
\end{eqnarray}
In the same spirit, for the exponential potential,
\begin{equation}\label{eq:exponential potential}
V_\textrm{exp}(\phi) = V_1 \exp \left[ - \phi / M_1 \right]  \, ,
\end{equation}
we identify \textit{$M_1$} as the characteristic constant parameter and accordingly define the characteristic $Q_\textrm{exp}(z)$,
\begin{eqnarray}\label{eq:Qexp}
 Q_\textrm{exp} (z)
  & \equiv & - V(z) \left( \frac{dV}{d\phi} \right)^{-1} (z)\label{eq:Qexp1} \\
  & = & M_1\quad \textrm{for the exponential potential.}\label{eq:Qexp2}
\end{eqnarray}
For the power-law potential,
\begin{equation}\label{eq:power-law potential}
V_\textrm{power}(\phi) = m^{4-n} \phi^n  \, ,
\end{equation}
we define the following characteristic corresponding to the index $n$,
\begin{eqnarray}
 Q_\textrm{power}(z)
   & \equiv & \left[ 1 - V(z) \left(\frac{dV}{d\phi}(z) \right)^{-2}
              \frac{d^2 V}{d\phi^2} (z) \right]^{-1} \label{eq:Qpower1} \\
   & = & n \quad \textrm{for the power-law potential.}\label{eq:Qpower2}
\end{eqnarray}
For the inverse-exponential potential,
\begin{equation}\label{eq:inverse-exponential potential}
V_\textrm{inverse-exp}(\phi) = V_2 \exp \left[ M_2 / \phi   \right]  \,,
\end{equation}
the characteristic is defined as
\begin{eqnarray}
 Q_\textrm{inverse-exp}(z)
   & \equiv & - \frac{4}{V(z)} \left(\frac{dV}{d\phi}(z) \right)^{3}\left[\frac{d^2 V}{d\phi^2} (z) - \frac{1}{V(z)}\left(\frac{dV}{d\phi}(z) \right)^{2} \right]^{-2}\label{eq:Qinverse-exp1} \\
   & = & M_2 \quad \textrm{for the inverse-exponential potential.}\label{eq:Qinverse-exp2}
\end{eqnarray}
As the dark energy component, the generalized Chaplygin gas has an equation of state govern by
\begin{equation}\label{eq:Chaplygin}
p_\textsc{de}(z) = -A / \left[\rho_\textsc{de}(z)\right]^{\alpha} \, ,
\end{equation}
where $\alpha\neq-1$ and $A>0$. The corresponding characteristic is defined as
\begin{eqnarray}
Q_\textrm{Chaplygin} (z)
  & \equiv & -   \frac{\rho_\textsc{de}(z)}{w(z)}  \frac{dw}{dz}(z)\left( \frac{d\rho_\textsc{de}}{dz}(z) \right)^{-1} -1\label{eq:QChaplygin1} \\
  & = & \alpha \quad \textrm{for the generalized Chaplygin gas.}\label{eq:QChaplygin2}
\end{eqnarray}

We then define the measure of consistency $\mathcal{M}_i(z)$ as the derivative of the characteristic $Q_i(z)$ with respect to the redshift for each dark energy model,
\begin{eqnarray}\label{eq:Mi}
 \mathcal{M}_i(z)
   & \equiv & \frac{dQ_i}{dz}(z) \\
   & = & 0 \quad \textrm{for the corresponding dark energy model,}
\end{eqnarray}
where $i$ denotes ``$\Lambda$'', ``exp'', ``power'', ``inverse\textrm{--}exp'' and ``Chaplygin'', respectively. $\mathcal{M}_i(z)$ can in general evolve with the redshift but should be constant zero in the domain of the corresponding dark energy model.

\subsection{Observational data and constraint}
\label{sec:II-B}


We use the combined data set from three types of observations
including the SN Ia observation, the CMB measurement, and the BAO
measurement.

We use the Constitution set of SN Ia data compiled by Hicken et
al.,\cite{Hicken:2009dk}$,$\cite{Riess:2004nr}$\textrm{--}$\cite{Kowalski:2008ez}
which provides the information of the luminosity distance and the
redshift. The luminosity distance-redshift relation is given by
\begin{equation} \label{eq:DL}
d_L(z)=(1+z)\int_0^{z}\frac{dz'}{H(z')}.
\end{equation}
We use the CMB shift parameter measured by the five-year WMAP
observation,\cite{Komatsu:2008hk}
\begin{equation} \label{eq:R}
 R = \sqrt{\Omega _m H_0^2} \int_0^{1090.04}\frac{dz}{H(z)}=1.710\pm 0.019.
 \end{equation}
We use the BAO measurement from the joint analysis of the SDSS and
$2$dFGRS data,\cite{Percival:2007yw,Eisenstein:2005su} which gives

\begin{equation}\label{eq:BAO}
D_V(0.35)/D_V(0.2) = 1.812\pm 0.060,
\end{equation}
where
\begin{equation}\label{eq:BAO1}
D_V(z_\textsc{bao}) = \left[(1+z_\textsc{bao})^2
D_A^2(z_\textsc{bao})
\frac{z_\textsc{bao}}{H(z_\textsc{bao})}\right]^{1/3},
\end{equation}
and $D_A(z)$ is the angular diameter distance,
\begin{equation} \label{eq:DA}
D_A(z)=\frac{1}{1+z}\int_0^{z}\frac{dz'}{H(z')}.
\end{equation}

\begin{figure}[ph]
\includegraphics[width=1\linewidth]{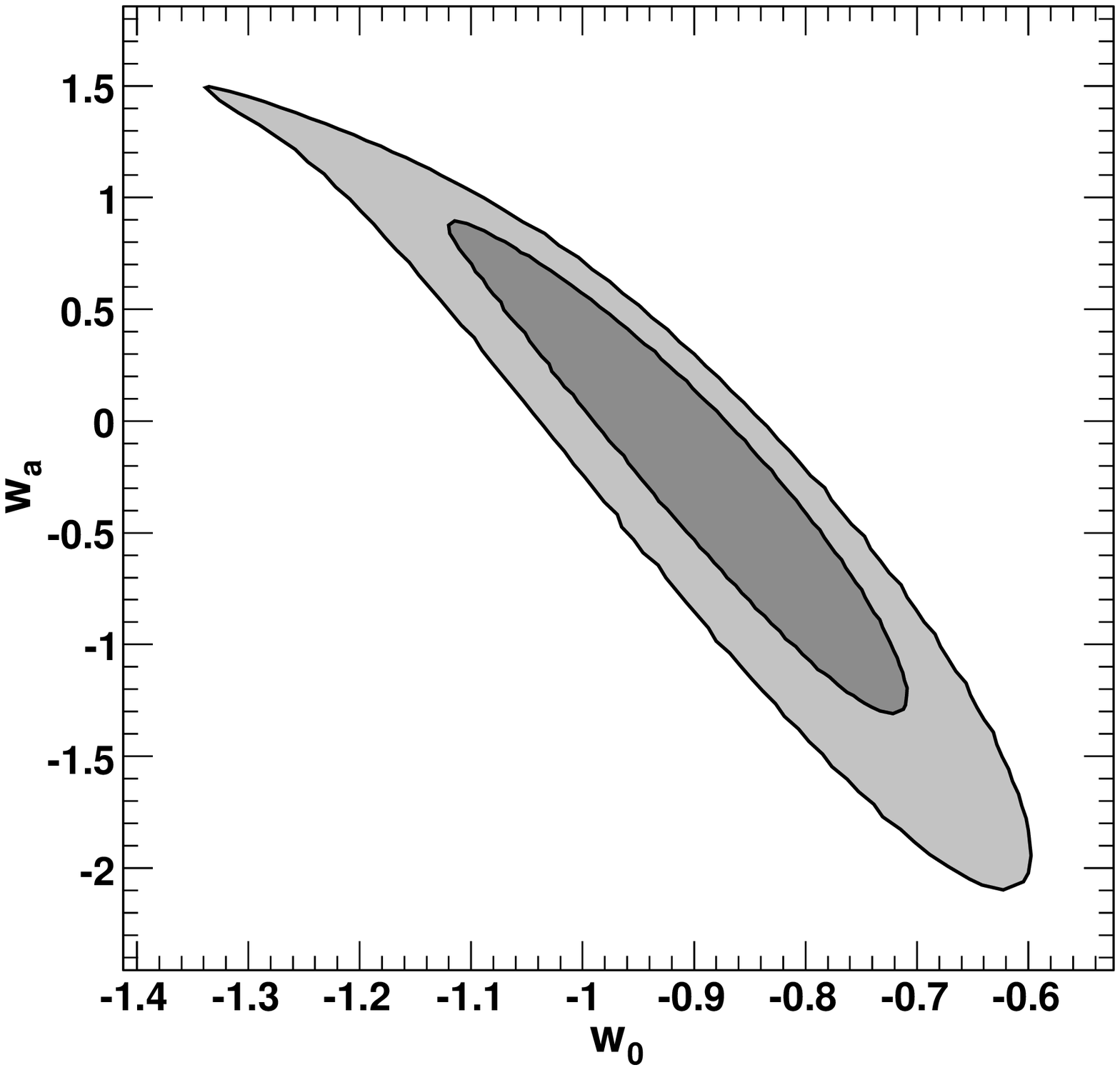}
\caption{The joint two-dimensional constraint of $(w_0, w_a)$ based
on a combined data set including the Constitution set of SN Ia data,
the CMB measurement from the five-year WMAP, and the BAO measurement
from the SDSS and $2$dFGRS. The dark and light gray areas correspond
to the $68.3\%$ and $95.4\%$ confidence regions, respectively.
\protect\label{fig1}}
\end{figure}

The constraint of the parameter space ($w_0$, $w_a$, $\Omega_m$) is
obtained by fitting the three parameters to this combined data set.
The best fit of the parameters are found to be
$$  w_0=-0.89^{+0.12}_{-0.14}, \quad w_a=-0.18^{+0.71}_{-0.74}, \quad \Omega_m= 0.25^{+0.03}_{-0.02}. $$
The two-dimensional constraint of $(w_0, w_a)$ is shown in
Fig.~\ref{fig1}.

\subsection{Results of the consistency test}
\begin{figure}[ph]
\includegraphics[width=1\linewidth]{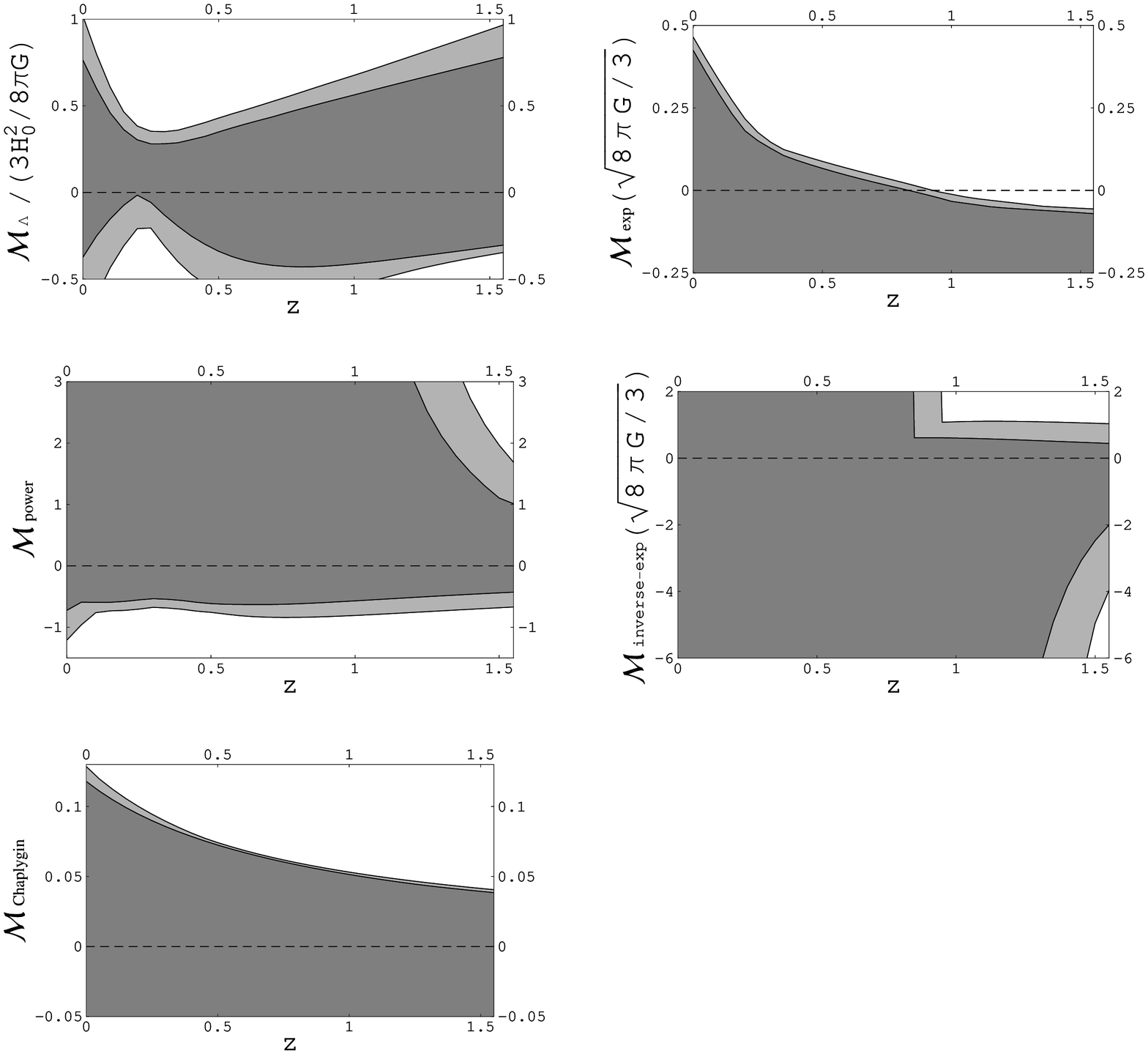}
\caption{The measure of consistency of the five dark energy models.
The dark and light gray areas correspond to the $68.3\%$ and
$95.4\%$ confidence regions, respectively. The
$\mathcal{M}_\textrm{exp}(z)=0$ line lies outside the $95.4\%$
confidence region for $0.95<z<1.55$. The null lines of the measure
for the other four models lie inside the $68.3\%$ confidence regions
for $0<z<1.55$. This indicates that the exponential potential is
ruled out at the $95.4\%$ confidence level while the other four dark
energy models are still consistent with the current observational
constraints down to the $68.3\%$ confidence
level.\protect\label{fig2}}
\end{figure}

For the consistency test of each dark energy model, we reconstruct
$\mathcal{M}_i(z)$ via the constraint of ($w_0$, $w_a$, $\Omega_m$),
with the use of the equations in Sec.~\ref{sec:introduction} and
Sec.~\ref{sec:Formalism}. We perform the test in the redshift region
$0<z<1.55$, where the influence of dark energy on the expansion of
the universe is most significant. This region is covered by the
current SN Ia observations, which is the most sensitive type of
observations to probe the behavior of dark energy. If the
$\mathcal{M}_i(z) = 0$ line lies outside certain confidence region,
the corresponding dark energy model is ruled out at that confidence
level. Adopting the constraint obtained in Sec.~\ref{sec:II-B}, we
find that the $\mathcal{M}_\textrm{exp}(z)=0$ line lies outside the
$95.4\%$ confidence region while the null lines of the other four
models lie inside the $68.3\%$ confidence region. The results are
shown in Fig.~\ref{fig2}.

\section{Summary\label{sec:summary}}

We have preformed consistency test of five dark energy models, including the cosmological constant, the generalized Chaplyngin gas, and three quintessence models: exponential, power-law, and inverse-exponential potentials. This test gives a simple signature if a dark energy model is ruled out by the observational data. It can be done efficiently via the constraint of a single set of parameters deduced from the observational data, and can test quintessence models without solving the field equation.

Through our approach and invoking the broadly used parametrization of the equation of state, the exponential potential is found to be ruled out at the $95.4\%$ confidence level based on the current observational data. The other four dark energy models remain consistent with the current observations down to the $68.3\%$ confidence level. It is worth noticing that in our previous work the power-law potential was ruled out at the $68.3\%$ confidence level based on a different data set \cite{Gu:2008ch}. One noticeable change in the new data set is that the cosmological constant is contained in the $68.3\%$ confidence region, which was not so in the previous one. Whether our method can discriminate between the power-law potential and the cosmological constant can in principle be studied with the Monte Carlo test \cite{Chen:2009}. The flat cosmological constant model was examined by Davis et al \cite{Davis:2007na}. via the model-based approach, in which they used the same data set except the three-year WMAP data instead. They found the goodness of fit for the flat cosmological constant model to be $43.7\%$ while we find the model at least consistent with the data at the $68.3\%$ confidence level. The two results are not in conflict with each other.

Our method of the consistency test can in principle be applied not only to other dark energy models but also to other models explaining the acceleration of the expansion, as long as one looks for the characteristic $Q(z)$ corresponding to a constant parameter of each model. One can also choose a different parametrization for better discriminating power between the models in regard. The discriminating power of the method with different forms of parametrization and the possible bias imposed by the chosen parametrization should be studied via the Monte Carlo test \cite{Chen:2009}.

\begin{acknowledgments}
C.-W.~Chen is supported by the Taiwan National Science Council under
Project No. NSC 95-2119-M-002-034 and NSC 96-2112-M-002-023-MY3, and
P. Chen by Taiwan National Science Council under Project No. NSC
97-2112-M-002-026-MY3 and by US Department of Energy under Contract
No. DE-AC03-76SF00515.
\end{acknowledgments}

\end{document}